\documentclass[12pt,a4paper]{article}
\usepackage{graphicx}
\usepackage{times}
\title{\bf 3-D Radiative Transfer Modeling of Structured Winds in Massive Hot Stars with Wind3D}
\author{A. Lobel$^1$, J. A. Toal\'{a}$^2$, and R. Blomme$^1$\\
\vspace{1cm}\\
\normalsize $^1$ Royal Observatory of Belgium, Ringlaan 3, B-1180, Brussels, Belgium\\ 
\normalsize $^2$ Universidad Nacional Aut\'{o}noma de M\'{e}xico, Centro de Radioastronom\'{i}a y Astrof\'{i}sica,
\\ \normalsize Campus Morelia, Michoac\'{a}n, M\'{e}xico\\}
\date{\mbox{}}
\begin{document}
\maketitle
\pagestyle{empty}
%
% WE REDEFINE THE plain LaTeX PAGESTYLE !!! 
% THIS PAGESTYLE WILL BE USED FOR THE FIRST PAGE ONLY !
%
\def\bull{\vrule height .9ex width .8ex depth -.1ex}
\makeatletter
\def\ps@plain{\let\@mkboth\gobbletwo
\def\@oddhead{}\def\@oddfoot{\hfil\tiny\bull\quad
``The multi-wavelength view of hot, massive stars''; 39$^{\rm th}$ Li\`ege Int.\ Astroph.\ Coll., 12-16 July 2010 \quad\bull}%
\def\@evenhead{}\let\@evenfoot\@oddfoot}
\makeatother
%
% AND DEFINE OUR MACROS FOR THE REFERENCE LIST
% I.E \beginrefer \refer and \endrefer
%
\def\beginrefer{\section*{References}%
\begin{quotation}\mbox{}\par}
\def\refer#1\par{{\setlength{\parindent}{-\leftmargin}\indent#1\par}}
\def\endrefer{\end{quotation}}
%
% BEGIN THE ABSTRACT CHAPTER WITH \noindent\small, ENCLOSE IT IN A GROUP
% AND BOLDFACE THE TITLE.
%
{\noindent\small{\bf Abstract:}
We develop 3-D models of the structured winds of massive hot stars
with the {\sc Wind3D} radiative transfer (RT) code. We investigate the 
physical properties of large-scale structures observed 
in the wind of the B-type supergiant HD~64760 with 
detailed line profile fits to Discrete Absorption 
Components (DACs) and rotational modulations observed with 
IUE in Si~{\sc iv} $\lambda$1395. 

We develop parameterized input models for {\sc Wind3D} with
large-scale equatorial wind density- and velocity-structures,
or so-called `Co-rotating Interaction Regions' (CIRs) 
and `Rotational Modulation Regions' (RMRs). The parameterized 
models offer important advantages for high-performance RT 
calculations over ab-initio hydrodynamic input models. 
The acceleration of the input model calculations permits 
us to simulate and investigate a wide variety of physical 
conditions in the extended winds of massive hot stars.

The new modeling method is very flexible for
constraining the dynamic and geometric wind properties 
of RMRs in HD~64760. We compute that the modulations are 
produced by a regular pattern of radial density 
enhancements that protrude almost linearly into
the equatorial wind. We find that the modulations are 
caused by narrow `spoke-like' wind regions. 
We  present a hydrodynamic model showing that the 
linearly shaped radial wind pattern can be caused 
by mechanical wave action at the base of the stellar 
wind from the blue supergiant.
}
%
% NOW COMES THE MAIN BODY OF THE ARTICLE
%
\section{Introduction}
Accurate mass-loss rates of massive hot stars determined from 
quantitative spectroscopy are important for understanding 
the physical properties of the radiative wind driving mechanism 
that is influenced by dynamic structures on both large and small 
length scales in the wind. Rotational modulations and Discrete 
Absorption Components are important tracers of large-scale structures 
in the highly supersonic winds of these stars. DACs are 
recurring absorption features observed in UV resonance lines 
of many OB-stars. They drift bluewards in the absorption 
portion of P-Cygni profiles. DACs are caused by spiral-shaped 
density- and velocity structures winding up in the plane of 
the equator over several tens of stellar radii (e.g., Cranmer 
\& Owocki 1996). Lobel \& Blomme (2008) demonstrated with 
3-D RT modeling (combined with hydrodynamic simulations) 
of the detailed DAC evolution in HD~64760 (B0.5 Ib) that 
these wind spirals are extended density waves emerging from 
two bright equatorial spots that rotate five times slower 
than the stellar surface. Detailed hydrodynamic models of 
its structured wind with these large density waves (or CIRs) 
reveal only a very small increase of less than 1~\% above 
the smooth (symmetric) wind mass-loss rate.

\begin{figure}[t]
\begin{minipage}{8cm}
\centering
\includegraphics[width=8cm]{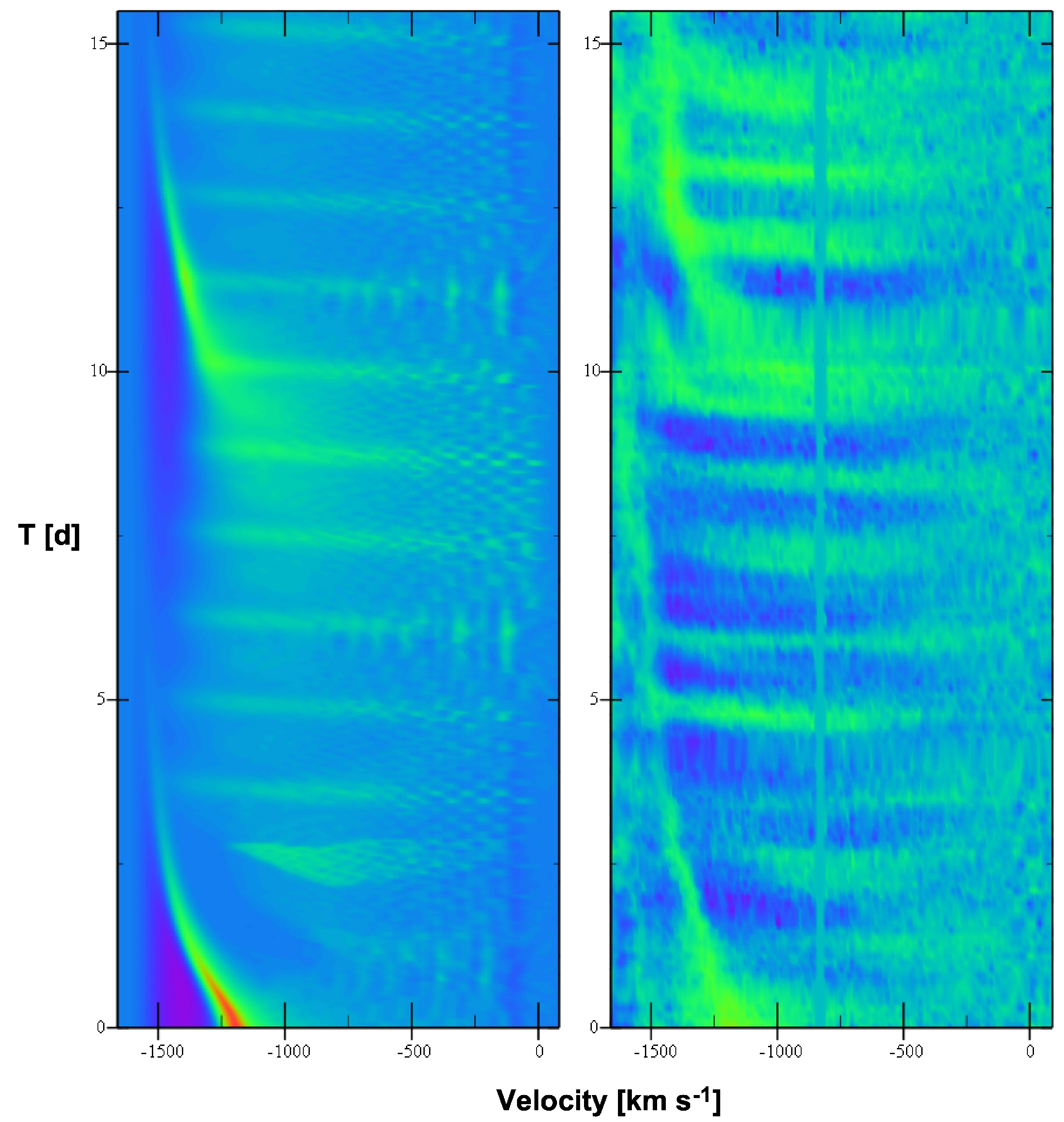}
\caption{Dynamic spectrum of Si~{\sc iv} $\lambda$1395 observed during 
15.5 d in HD~64760 ({\it right-hand panel}), compared to 3-D 
radiative transfer calculations ({\it left-hand panel}) with a 
parameterized structured wind model. Horizontal absorptions are 
the modulations we model in this paper. \label{fig_1}}
\end{minipage}
\hfill
\begin{minipage}{8cm}
\centering
\includegraphics[width=8cm,height=8cm]{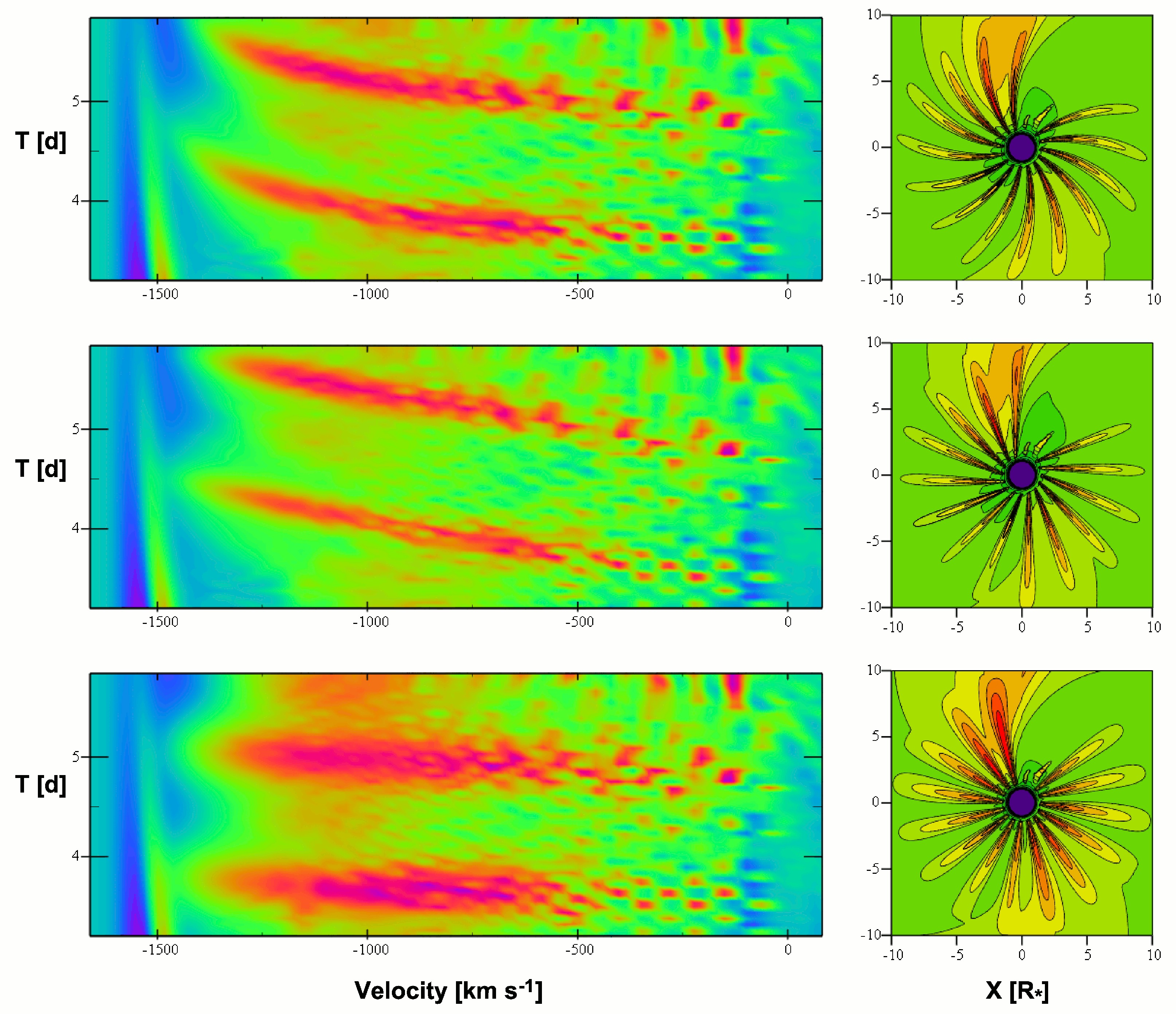}
\caption{The curvature, incidence angle on the 
surface, and opening angle of the RMRs shown in the right-hand panels 
determine the acceleration ({\em upper panels}), inclination 
({\em middle panels}), and duration ({\em lower panels}), 
respectively, of the RT calculated modulations ({\em left-hand panels}).
\label{fig_2}}
\end{minipage}
\end{figure}

The bright surface spots produce the large-scale CIRs in the 
wind. The density enhancements and velocity plateaus of 
the CIRs yield migrating DACs with a recurrence 
period of 10.3 d in the UV line profiles of HD~64760. 
The modulations, on the other hand, show much shorter 
periods of $\sim$1.2~d and reveal a time-evolution 
that substantially differs from the rather slowly 
shifting DACs. The modulations are nearly-flat 
absorption features observed for only $\sim$0.5~d
to 0.75~d with radial velocities that can range from 
$\sim$0 $\rm km\,s^{-1}$ to $\sim$$v_{\infty}$=1560 
$\rm km\,s^{-1}$ (Massa et al. 1995; Prinja 1998). 
They can intersect the slowly drifting DACs and
sometimes reveal a remarkable bow shape (Fullerton 
et al. 2006) with broad flux minima around 
$\sim$930~$\rm km\,s^{-1}$. In this paper we 
present a semi-empiric
model of the large-scale wind structures based 
on detailed RT fits to the time-evolution 
of the modulations observed in HD~64760. 
We utilize the {\sc wind3d} code for 
performing non-LTE RT calculations ({\em Sect.~2}) 
of important diagnostic spectral lines, such as 
Si~{\sc iv} $\lambda$1395. We develop 
{\em parameterized} 3-D input models for Wind3D 
in Sect. 3 because they offer important advantages for 
high-performance RT calculations over 
ab-initio hydrodynamic input models. 
The acceleration of the input model 
calculations permits us to model and investigate 
a much broader range of 3-D physical conditions in the 
wind. 

\section{Parametrization of Wind3D Input Models}
Wind3D computes the 3-D non-LTE transport of radiation 
for the 2-level atom in optically thick resonance 
lines formed in scattering-dominated extended stellar winds. 
It lambda-iterates the line source function 
on a Cartesian grid of $71^{3}$ points. 
The radiative transfer equation is solved in parallel 
over a mesh of $701^{3}$ voxels using the converged 
line source function. Typical computation times with 
{\sc wind3d} are $\sim$5 h for iterating the 
line source function, and $\sim$12 h for 
calculating the dynamic line profile variability 
at 90 viewing angles using 16 CPUs.

We develop a new software module in {\sc wind3d} 
for semi-empiric modeling of large-scale equatorial wind 
density- and velocity-structures. The computer code 
integrates the time-independent momentum balance equation 
of radiation-driven rotating winds following Castor, 
Abbott \& Klein (CAK, 1975). 
First, we parametrize the large-scale density structures in a
stationary model of the equatorial wind with:
\begin{equation}
\rho(x,y,\psi) = \rho_{0}(r)\,\left( 1 + A(r)\,{\rm sin}^{m}\left(\psi\,n+2\pi\,
n\,f_{c}\left(\frac{{\rm R_{\star}}-\sqrt{(x-x_{0})^{2}+y^{2}}}{{\rm R_{\star}}-{\rm R_{mod}}}
\right)\right) \right) \,,
\end{equation}
where $r^{2}$=$x^{2}$+$y^{2}$, $\rm R_{cr}$ $\leq r \leq$ $\rm R_{mod}$, 
$-\pi$ $\leq \psi$ $\leq$ $\pi$, and $\rm R_{mod}$ is the outer 
radius of the wind model. $x_{0}$=$\rm R_{\star}$$\rm sin$$(\psi_{0})$, where
$\psi_{0}$ is the incidence angle of the equatorial wind structures on 
the stellar surface, and 0 $\leq$ $\psi_{0}$ $\leq$ $\pi$/2.
$\rho_{0}(r)$ is the density of the smooth (unperturbed) equatorial 
wind. $n$ denotes the number of large-scale wind structures with 
enhanced density in one hemisphere, and $m$ is an even exponent of the 
sine function that sets the opening angle or the tangential width 
of the wind structures. In wind regions where the sine function vanishes 
the local wind density equals the smooth wind density. 
The term in Eq. (1) with $f_{c}$, where $0 \leq$ $f_{c}$ $\leq$ 1, determines the 
curvature of the wind structures between $\rm R_{\star}$ and $\rm R_{mod}$. If 
$f_{c}$=1, they turn completely around the star over 2\,$\pi$ from the surface
radius $\rm R_{\star}$ to $\rm R_{mod}$. In case $f_{c}$=0, the structures do
not curve at all and stay strictly radial (linear) in the wind, 
ordered in equal sectors around the star. 
We parametrize the density constrast of the large-scale 
wind structures with the function $A$ in Eq. (1):
\begin{equation}
A(r) = c_{1}\,{\rm exp} \left( -\left(\frac{r}{{\rm R_{\star}}}-c_{2}
\right)^{2} +c_{3}\,\left(\frac{r}{{\rm R_{\star}}}\right)^{2}+c_{4}  
\right) \,,
\end{equation}    
where $c_{1,2,3,4}$ are constants. They determine the detailed density 
profile compared to $\rho_{0}(r)$. We vary the four constants until the 
parameterized density profile best fits the density contrast of 
the CIRs in the hydrodynamic wind model of HD~64760.

Next, with the parameterized radial wind density-structure we
compute the corresponding radial wind velocity-structure for
the $\alpha$ parameter of CAK theory set equal to 1/2 in 
hot stars with 20 kK $\leq$ $T_{\rm eff}$ $\leq$ 30 kK.
We solve the CAK momentum balance equation in 1-D for many
angles $\psi$, and include the centrifugal force in the 
equatorial plane due to the photospheric rotation 
velocity $v_{\rm rot}$. With the conservation of angular 
momentum the equatorial tangential wind velocity is      
$v_{\rm rot} {\rm R_{\star}}/{r}$, which becomes 
negligibly small in the highly supersonic outer regions 
of the parameterized wind models. We integrate the momentum
equation together with the mass continuity equation towards the 
surface from a typical outer (reference) radius 
$r_{\rm ref}\geq$30~$\rm R_{\star}$ with the boundary condition 
$v(r=r_{\rm ref})=v_{\infty}$, down to the critical point $\rm R_{cr}$
of the wind. The value of the mass-loss rate for calculating the CAK 
line force is computed with Eq. (5) of Lobel \& Blomme (2008), but 
which typically increases the structured wind mass-loss rate by 
less than $\sim$1~\% in the best-fit 2-spot model. We find that 
the radial wind velocity-structure computed with the CAK 
wind-momentum equation integration method, using parameterized 
stationary wind density models, compares very closely to the hydrodynamic 
wind velocity model of HD~64760 (Lobel \& Toal\'{a} 2009). 

\section{Radiative Transfer Modeling of Rotational Modulations}

Figure 1 compares the dynamic spectrum of Si~{\sc iv}  
observed over 15.5~d in HD~64760 ({\em right-hand panel}) 
and computed with {\sc wind3d} ({\em left-hand panel}) 
using a parameterized structured wind model. The slowly 
bluewards migrating upper and lower DACs result from two 
spiral-shaped CIRs shown in the right-hand panels of Fig.~2. 
The nearly horizontal absorption in the modulations is
due to almost linearly shaped density enhancements and 
local wind velocity variations that radially 
protrude into the equatorial wind. The RMRs have 
maximum tangential widths of less than $\sim$1~$\rm R_{\star}$ 
across the wind. Our parameterized RT modeling reveals that the 
RMRs are {\em linear} density enhancements through the wind 
because the modulations of HD~64760 stay flat beyond 
1000~$\rm km\,s^{-1}$. The upper panels of Fig. 2 show 
that the (radial velocity) acceleration of the modulations 
is too slow ({\em left-hand panel}) in case the RMRs 
curve too quickly over $\sim$10~$\rm R_{\star}$ above the 
stellar surface ({\em right-hand panel}). Hence, the 
RMRs are large-scale density- and velocity-structures 
in the supersonically (radiatively) accelerating equatorial 
wind. The parameterized best-fit modeling method shows that 
the RMRs are centered around the star with small inclination 
angles of $\leq 6^{\circ}$ from the radial direction 
(the rather narrow RMRs have small incidence angles on 
the stellar surface in the middle panel of Fig.~2). 
We also compute opening angles of $\leq 10^{\circ}$ 
for the RMRs at the wind base to correctly fit 
the observed duration of the modulations. 

\begin{figure}[t]
\begin{minipage}{8cm}
\centering
\includegraphics[width=8cm,height=8cm]{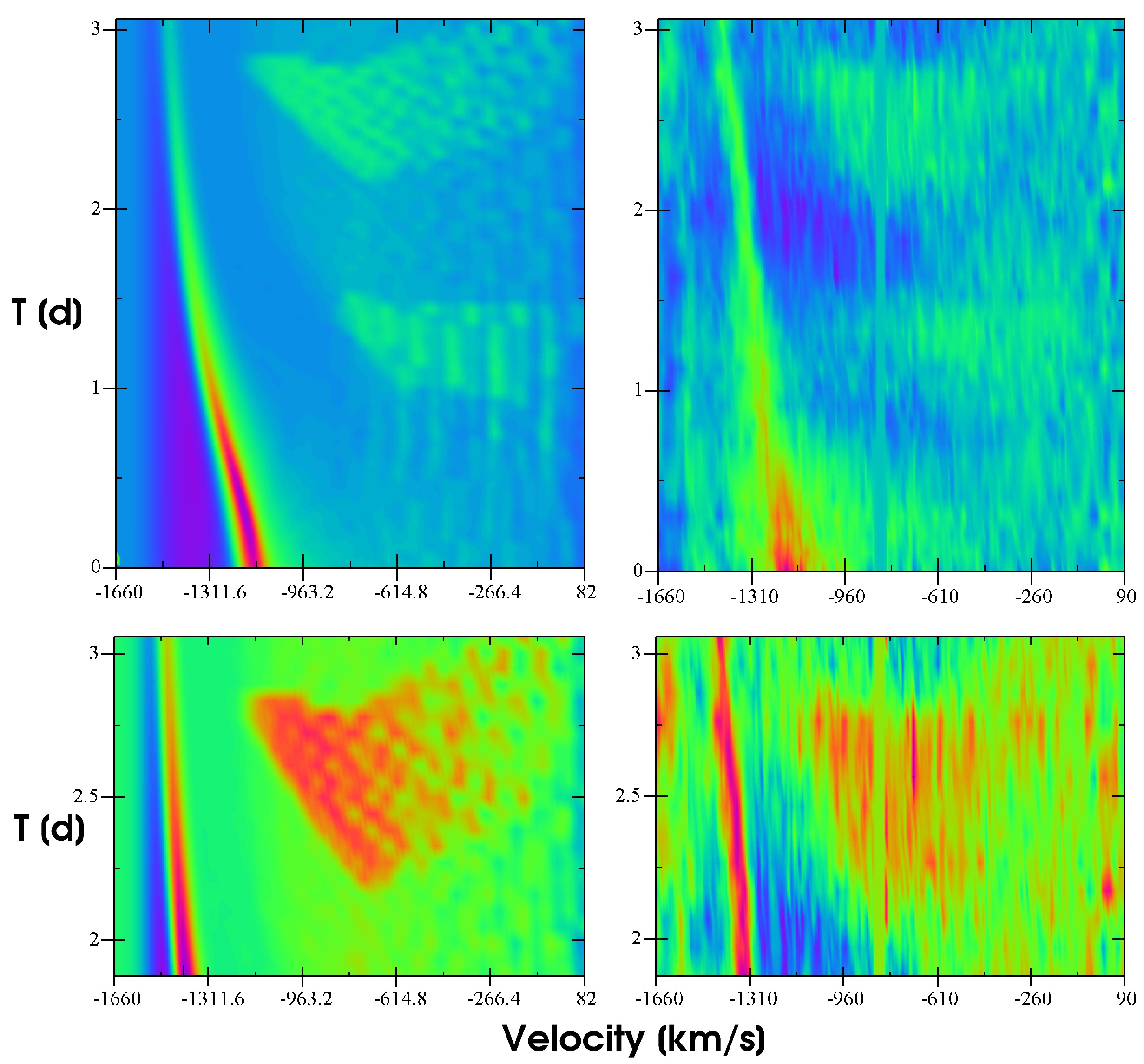}
\caption{The upper panels show two modulations computed with 
Wind3D ({\em left-hand panel}) and observed over $\sim$3.1~d 
({\em right-hand panel}) in Si~{\sc iv} $\lambda$1395 of HD~64760. 
The lower panels show the best fit ({\em left}) and the observed
({\em right}) upper `bow shaped' modulation in more detail 
({\em see text}). \label{fig_3}}
\end{minipage}
\hfill
\centering
\begin{minipage}{8cm}
\includegraphics[width=8cm]{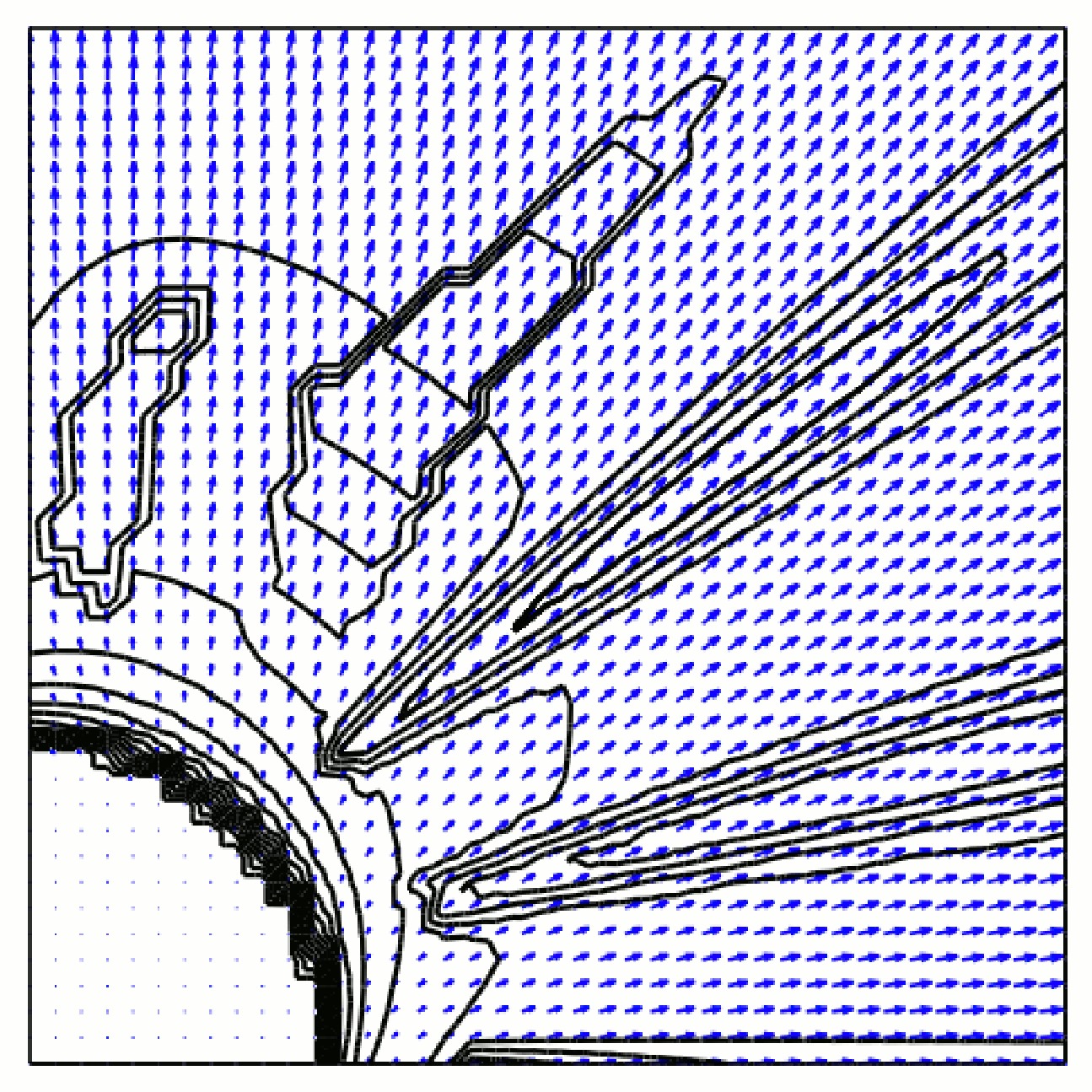}
\caption{Parameterized model of the equatorial wind-density 
structure of HD~64760 computed with 3-D RT fits to the 
modulations in Fig.~3. The solid drawn 
lines show density contours of the modulations. 
The blue arrows mark the local wind 
velocity to $\sim$3~$\rm R_{\star}$ above the stellar surface. 
\label{fig_4}}
\end{minipage}
\end{figure}

Figure~3 shows a portion of the dynamic spectrum in 
Fig.~1 observed between 0~d and 3.1~d ({\it upper 
right-hand panel}). The lower DAC in Fig.~3 slowly 
shifts bluewards from $\sim$1100~$\rm km\,s^{-1}$ 
to $\sim$1400~$\rm km \,s^{-1}$, while two modulations 
are observed around 1.2~d ({\em lower modulation}) and 
2.5~d ({\em upper modulation}). The upper modulation 
is observed during $\sim$0.7~d and reveals a somewhat curved 
shape, whereas the lower one occurs during 
$\sim$0.5~d showing a more irregular absorption 
pattern below $\sim$800~$\rm km\, s^{-1}$.
We delineate the borders of the wind density 
enhancements in two RMRs above the stellar surface 
and compute the radial wind velocity structure 
of the parameterized model by integrating 
the CAK momentum equation ({\em Sect. 2}). 
The lower panels of Fig.~3 show the observed 
({\em right-hand panel}) and best-fit 
({\em left-hand panel}) dynamic flux profile of 
the upper modulation computed with {\sc wind3d}. 
We fit the peculiar wedge-like 
shape at its short-wavelength side in detail 
by decreasing the opening angle of the density 
enhancement borders of the RMR beyond 
1~$\rm R_{\star}$ above the stellar 
surface, shown in Fig.~4. The radial wind 
velocity-structure of the upper modulation does not 
exceed $\sim$1200~$\rm km\,s^{-1}$ around 2.8~d, 
and it does not intersect the velocity of the lower DAC 
around $\sim$1400~$\rm km\,s^{-1}$. The wind density 
and velocity model of the upper modulation in Fig.~4 
does therefore not exceed a distance of 
$\sim$2.5~$\rm R_{\star}$ above the stellar surface
because the smooth-wind velocity exceeds 
$\sim$1200~$\rm km\,s^{-1}$ only beyond that radius. 
We therefore attribute the remarkable bow shape 
(called `phase-bowing') observed in the upper 
modulation to the intrinsically curved shape of 
the front and backside borders of the 
RMR we model in detail in Fig.~4. Several 
modulations observed between 5.5~d and 9~d in 
Fig. 1 do not show a clear bow shape (they 
are in fact strictly flat), although their flux 
profiles compare closely to the modulations 
that show the bow shape in Fig. 3. The remarkable 
bow shape is rather peculiar and results from the 
slightly curved shapes of a number of 
RMRs in the model. Modulations without the bow shape are
due to RMRs that protrude strictly radially 
into the wind. Best fits to the detailed flux profiles 
in the modulations show that the maximum density 
of the RMRs does not exceed the smooth-wind density 
by $\sim$17~\%, or about half the maximum density 
contrast of $\sim$31~\% in the CIR model of the 
lower DAC. Hence, the RMRs do not appreciably 
increase the wind mass-loss rate of HD~64760, 
also concluded for the CIR models in 
Lobel \& Blomme (2008).

\begin{figure}[t]
\begin{minipage}{8cm}
\centering
\includegraphics[width=8cm]{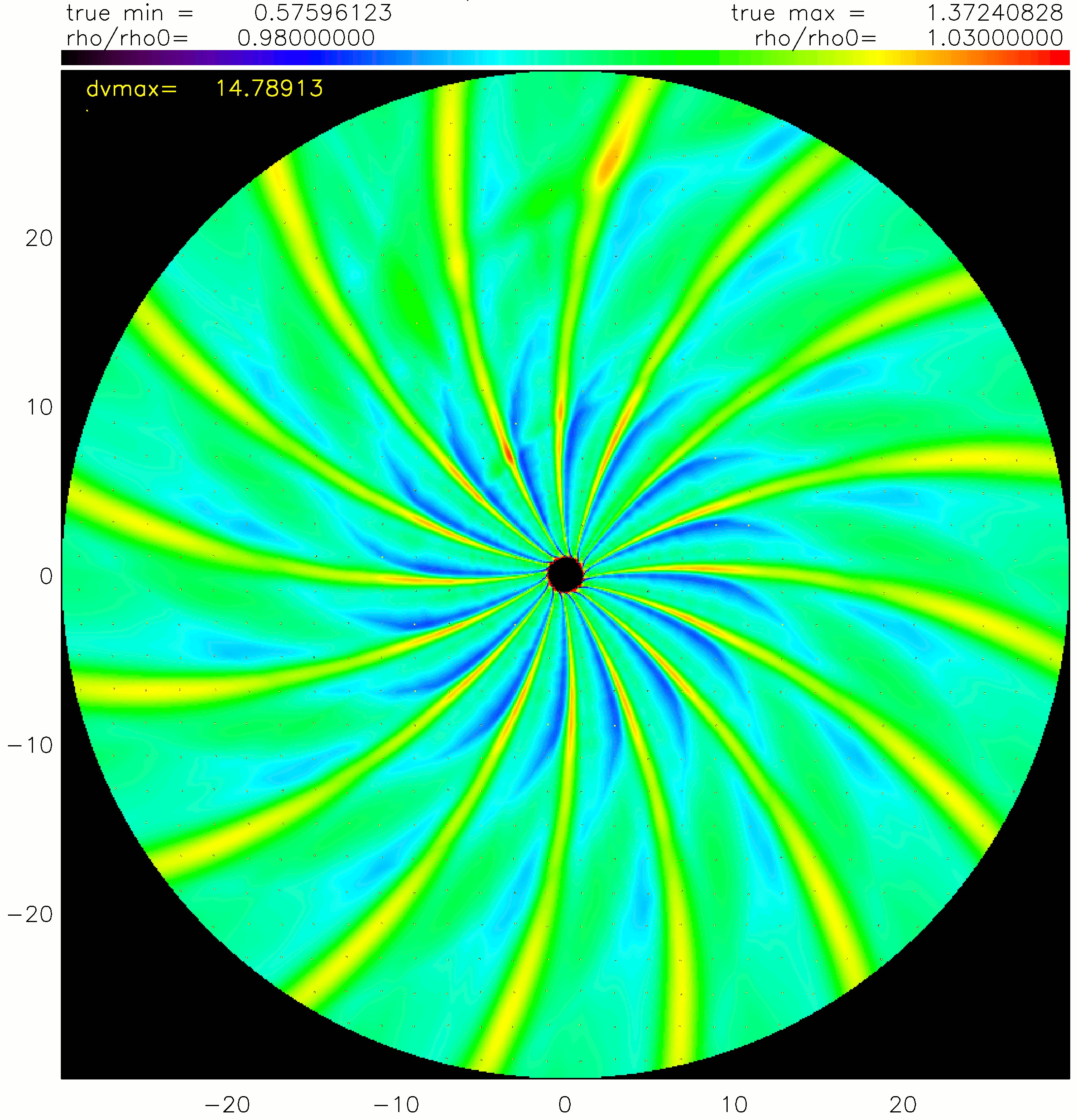}
\caption{Yellow colors mark density enhancements in 
the hydrodynamic wind model for rotational modulations 
of HD~64760 
({\em see text}). 
\label{fig_5}}
\end{minipage}
\hfill
\begin{minipage}{8cm}
\centering
\includegraphics[width=8cm]{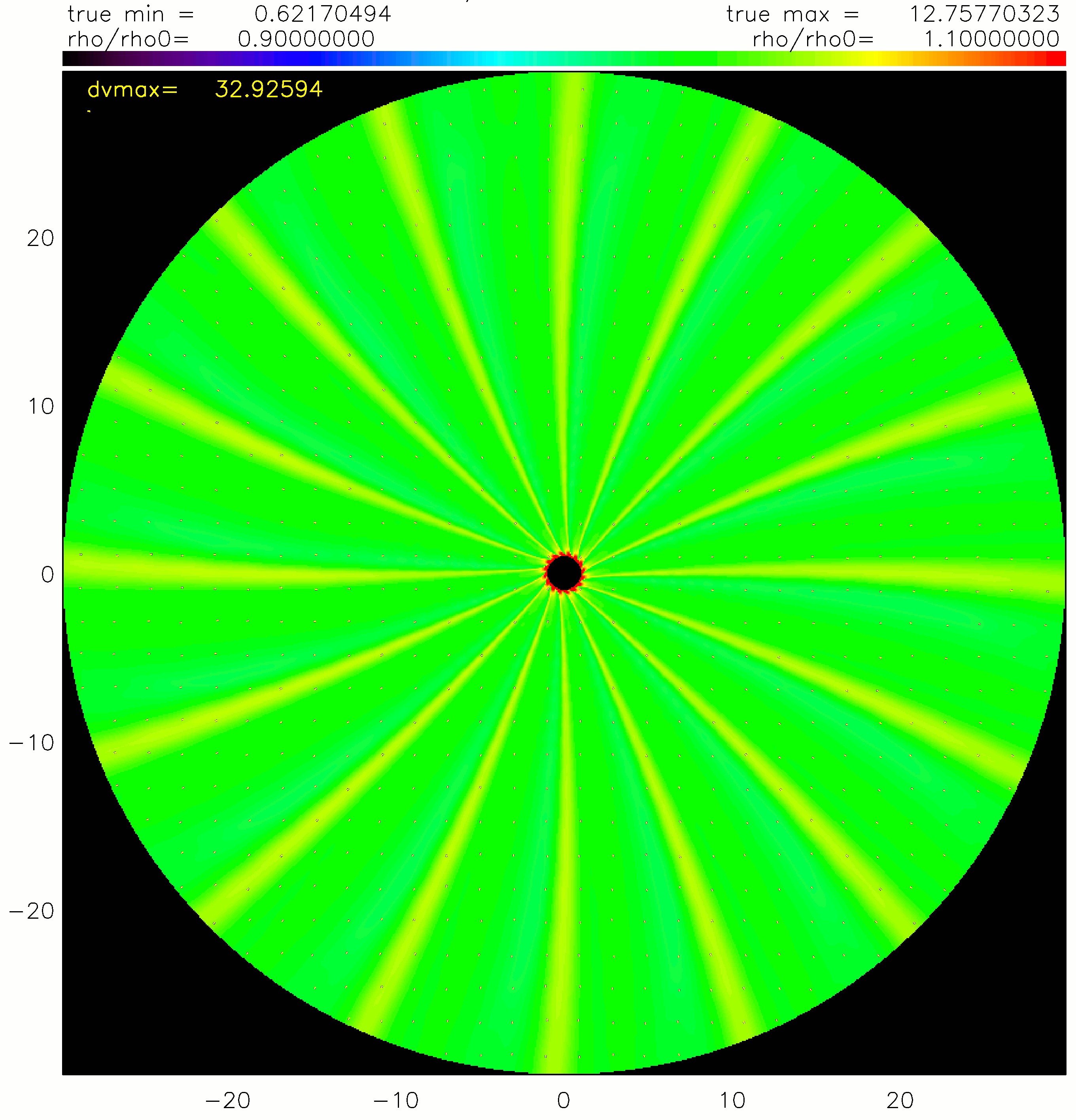}
\caption{The RMRs in the hydro model 
are less curved for a horizontal wave at the wind 
base that rotates 5 times slower than the star. \label{fig_6}}
\end{minipage}
\end{figure}

Owocki, Cranmer, \& Fullerton (1995) presented a kinematic model
for the spiral density wind features that satisfies the 
time-dependent mass-continuity relation. We further investigate 
the results of the (time-independent) semi-empiric modeling method 
with time-dependent hydrodynamic models. We compute hydroynamic 
test models of RMRs in HD~64760 with {\sc zeud3d}. 
We introduce a propagating pressure wave at the lower 
wind boundary near the stellar surface. The wave propagates 
horizontally over the stellar equator. It periodically 
alters the radiative wind acceleration due to
mechanical momentum imparted upwards to the wind by 
wave action. Figure 5 shows the density contrast computed 
in the wind out to 30~$\rm R_{\star}$ for a wave
with a radial velocity amplitude of $v_{\rm sound}$/100 
at the stellar surface. The wavelength is set to 1/16 of the 
stellar circumference and it co-rotates couter-clockwise
with the surface ($\omega_{\rm wave}$=$\omega_{\rm star}$). 
The density structure at the wind base is very intricate 
and converges into 16 rather narrow and stable density 
enhancements extending almost radially through the 
wind. The tangential width of the
wind density enhancements ({\em in yellow color}) 
is $\le 1$~$\rm R_{\star}$, comparable to the parameterized 
model we find for the RMRs. The hydrodynamic model in Fig. 6 
shows a decrease of the intrinsic curvature of these 
narrow wind features for a wave that rotates 5 times 
slower than the surface ($\omega_{\rm wave}$=$\omega_{\rm star}$ / 5).
The velocity amplitude of the wave is increased 
to $2 \times v_{\rm sound}$, producing somewhat 
larger densities of $\sim$8 \% above the smooth-wind 
density. The near-linear regular pattern in the 
hydrodynamic wind model is almost identical to the 
spoke-like RMRs we compute with parameterized models 
of the modulations in HD~64760. The maximum density contrast 
of the hydrodynamic wind pattern is however too 
small compared to $\sim$17~\% of the RMRs to fit 
the flux profiles observed in the modulations. 
Further hydrodynamic modeling is required 
to simulate the conditions of the best-fit parameterized 
wind model in more detail, and to investigate 
the formation physics of the large-scale wind pattern 
that causes the modulations observed in HD~64760. 

\section{Conclusions}
We perform 3-D RT calculations with {\sc wind3d} 
of the rotational modulations observed in Si~{\sc iv} 
$\lambda$1395 of HD~64760. We find that the
horizontal absorptions in the line are caused by 
a very regular pattern of almost linearly shaped radial 
density- and velocity-variations in the equatorial 
wind out to $\sim$10~$\rm R_{\star}$ above the stellar surface. 
The density in the RMRs does not exceed $\sim$17~\% of
the smooth-wind density, and hence they do not appreciably 
increase the stellar mass-loss rate. Hydrodynamic 
models computed with {\sc zeus3d} show that RMRs 
can result from mechanical wave action at the 
base of the stellar wind producing the rather narrow 
`spoke-like' wind regions in a regular geometric pattern 
centered around the star. We conjecture that the waves are caused by 
non-radial pulsations (for example discussed 
in Kaufer et al. 2006) of HD~64760.

%
% USE A SECTION WITHOUT NUMBER FOR THE ACKNOWLEDGEMENTS
%
%\section*{Acknowledgements}
%
%
% BEGIN THE REFERENCE LIST WITH \beginrefer
% USE \refer BEFORE THE REFERENCES AND BEGIN A NEW PARAGRAPH AFTER THE 
% REFERENCE !
% DO NOT FORGET TO END THE LIST WITH \endrefer
%
\footnotesize
\beginrefer

\refer Castor, J. I., Abbott, D. C. \& Klein, R. I. 1975, ApJ, 195, 157

\refer Cranmer, S. R., \& Owocki, S. P. 1996, ApJ, 462, 469

\refer Fullerton, A. W., Massa, D. L., Prinja, R. K., Owocki, S. P., 
\& Cranmer, S. R. 1997, A\&A, 327, 699

\refer Kaufer, A., Stahl, O., Prinja, R. K., \& Witherick, D. 2006, A\&A, 447, 325

\refer Lobel, A., \& Blomme, R. 2008, ApJ, 678, 408

\refer Lobel, A., \& Toal\'{a}, J. A. 2009, in {\em Eta Carinae in the Context
of the Most Massive Stars}, Proc. of XXVIIth IAU GA, 
Highlights of Astronomy 14, ed. I. F. Corbett, p. 20, arXiv:0910.3158v1 
 
\refer Massa, D., et al. 1995, ApJ, 452, L53

\refer Owocki, S. P., Cranmer. S. R., \& Fullerton, A. W. 1995, ApJ 453, L37

\refer Prinja, R. K. 1998, in {\em Cyclical Variability in Stellar Winds}, 
eds. L. Kaper \& A. W. Fullerton, 92, Springer-Verlag, Heidelberg

\endrefer           
\end{document}